\documentclass[a4paper,aps,prl,showpacs,preprintnumbers,twocolumn,nofootinbib,superscriptaddress]{revtex4-1}

\usepackage[english]{babel}

\usepackage[utf8x]{inputenc}
\usepackage[T1]{fontenc}
\usepackage{ucs}
\usepackage{microtype}
\usepackage{newtxtext,newtxmath}
\usepackage{color}

\usepackage{url}
\usepackage{hyperref}

\usepackage{xspace}
\usepackage{lastpage}
\usepackage{stmaryrd}

\usepackage{graphicx}
\usepackage{empheq}
\usepackage{ifthen}
\usepackage{tensor}
\usepackage{paralist}

\usepackage{amsmath}
\usepackage{amsfonts}
\usepackage{amssymb}
\usepackage{mathrsfs}
\usepackage{extarrows}
\usepackage{verbatim}
\usepackage{upgreek}
\usepackage{wasysym}

\usepackage{titlesec}

\usepackage{natbib}

\newcommand\ph{\ensuremath{\varphi}}
\newcommand\eps{\ensuremath{\varepsilon}}

\newcommand\define{\equiv}

\newcommand\vect[1]{\boldsymbol{#1}}

\newcommand\ex[1]{\mathrm{e}^{#1}}
\renewcommand\i{\ensuremath{\mathrm{i}}}

\newcommand\transpose[1]{#1^{\rm T}}

\newcommand\e[1]{_{\text{#1}}}

\newcommand{\dd}{\mathrm{d}}

\renewcommand\lim[2]{\underset{ #1 \rightarrow #2 }{ \mathrm{lim} } \,}

\newcommand{\delimiters}[4][]{
\ifthenelse{ \equal{#1}{1} }{  #2 #3 #4  }
					{ \ifthenelse{\equal{#1}{2}}{ \big#2 #3 \big#4 }
						{ \ifthenelse{\equal{#1}{3}}{ \Big#2 #3 \Big#4 }
							{ \ifthenelse{\equal{#1}{4}}{ \bigg#2 #3 \bigg#4 }
								{ \ifthenelse{\equal{#1}{5}}{ \Bigg#2 #3 \Bigg#4 }
									{ \left#2 #3 \right#4 }
								}
							}
						}
					}
													}

\newcommand{\pa}[2][]{\delimiters[#1]{(}{#2}{)}}
\newcommand{\pac}[2][]{\delimiters[#1]{[}{#2}{]}}

\newcommand{\abs}[2][]{\delimiters[#1]{|}{#2}{|}}
\newcommand{\ev}[2][]{\delimiters[#1]{\langle}{#2}{\rangle}}

\newcommand{\apjl}{Astrophys. J. Lett.}
\newcommand{\aap}{A{\&}A}

\newcommand{\mnras}{MNRAS}

\newcommand{\jcap}{JCAP}

\newcommand{\physrep}{Phys. Rep.}
\renewcommand{\prl}{Phys. Rev. Lett.}
\newcommand{\azh}{Astron. Zh.}

\newcommand{\source}{\mathcal{S}}
\newcommand{\image}{\mathcal{I}}
\newcommand{\quadrupole}{\mathcal{Q}}
\newcommand{\amplification}{\mathcal{A}}

\titleformat{\section}[runin]
{\it\bfseries}
{}{0pt}{}[.---]
\titlespacing{\section}
{0pt}{1.5ex plus .1ex minus .2ex}{0pt}

\begin{document}

\title{Weak Gravitational Lensing of Finite Beams}

\author{Pierre Fleury}
\email{pierre.fleury@unige.ch}
\affiliation{D\'{e}partment de Physique Th\'{e}orique, Universit\'{e} de Gen\`{e}ve,\\
24 quai Ernest-Ansermet, 1211 Gen\`{e}ve 4, Switzerland}

\author{Julien Larena}
\email{julien.larena@uct.ac.za}
\affiliation{Department of Mathematics and Applied Mathematics\\
University of Cape Town,
Rondebosch 7701, South Africa}

\author{Jean-Philippe Uzan}
\email{uzan@iap.fr}
\affiliation{
            Institut d'Astrophysique de Paris, CNRS UMR 7095, Universit\'e Pierre et Marie Curie - Paris VI, 98 bis Boulevard Arago, 75014 Paris, France \\
           Sorbonne Universit\'es, Institut Lagrange de Paris, 98 bis, Boulevard Arago, 75014 Paris, France.}

\begin{abstract}
The standard theory of weak gravitational lensing relies on the infinitesimal light beam approximation. In this context, images are distorted by convergence and shear, the respective sources of which unphysically depend on the resolution of the distribution of matter---the so-called Ricci-Weyl problem. In this Letter, we propose a strong-lensing-inspired formalism to describe the lensing of finite beams. We address the Ricci-Weyl problem by showing explicitly that convergence is caused by the matter enclosed by the beam, regardless of its distribution. Furthermore, shear turns out to be systematically enhanced by the finiteness of the beam. This implies, in particular, that the Kaiser-Squires relation between shear and convergence is violated, which could have profound consequences on the interpretation of weak-lensing surveys.
\end{abstract}

\date{\today}
\maketitle

\section{Introduction}

Almost one century after the first measurements of the bending of light by the Sun, gravitational lensing has become one of the major tools of astrophysics and cosmology. Although it is a direct consequence of Maxwell's electromagnetism in curved spacetime, gravitational lensing enjoys a rich phenomenology~\cite{2006glsw.conf.....M}. In our Galaxy, the amplification of the light curve of stars by microlensing reveals the presence of exoplanets and constrains the nature of dark matter; giant arcs and Einstein rings produced by strong lensing allow us to measure the mass of galaxies and clusters, while correlations in the observed shape of galaxies due to weak lensing give access to the large-scale distribution of matter in the Universe.

From a theoretical perspective, the nontrivial connection between those regimes---especially weak and strong---is not fully understood yet. Indeed, the orthogonality between the underlying approximations led to the development of distinct languages. The weak-lensing formalism is based on a fully relativistic approach due to Sachs~\cite{1961RSPSA.264..309S}, where the propagation of light beams is dictated by the Riemann curvature they locally experience. In this context, lensing effects are fully encapsulated in two distortion modes, namely convergence~$\kappa$ and shear~$\gamma$---to which we could also add rotation, at second order. Thus, by construction, this approach is unable to address the phenomenology of strong lensing, where distortions of the light beams, such as giant arcs, cannot be reduced to three numbers only. The reason for this failure is that the Sachs machinery relies on the geodesic deviation equation, and, thus, supposes that the light beam can be considered \emph{infinitesimal}. This assumption does not hold in the strong lensing regime.

Another, though related, issue with the Sachs formalism concerns the dichotomy between the Ricci and Weyl components of spacetime curvature. On the one hand, Ricci curvature is directly related to the local energy density via Einstein's equation, and is a source of convergence; on the other hand, Weyl curvature encodes the long-range tidal fields generated by massive bodies, and is a source of shear. This distinction, however, depends on the resolution at which the matter distribution is considered: a coarse-grained distribution of matter, seen as a continuous medium, would mostly produce Ricci curvature, while in the limit of a set of point masses, curvature is Weyl everywhere, apart from peaks of Ricci curvature located on the masses themselves. 

The so-called \emph{Ricci-Weyl problem} has been extensively discussed in the context of cosmology~\cite{1964SvA.....8...13Z, DashevskiiZeldovich1965, 1966SvA.....9..671D, 1965AZh....42..863D, 1966RSPSA.294..195B, 1967ApJ...150..737G, 1967ApJ...147...61G, 1969ApJ...155...89K, DR72, 1973ApJ...180L..31D, 1973PhDT........17D, 1974ApJ...189..167D, 1975ApJ...196..671R, 1976ApJ...208L...1W,1998PhRvD..58f3501H,2012MNRAS.426.1121C,2012JCAP...05..003B}, the underlying question being: to which extent can the propagation of a light beam, which mostly occurs in regions dominated by Weyl curvature, be addressed using the Friedmann-Lema\^{i}tre-Robertson-Walker geometry, whose curvature is exclusively Ricci? This question led to a number of works which greatly improved our understanding of light propagation in inhomogeneous cosmologies~\cite{2005ApJ...632..718K, 2013PhRvL.110b1301B, 2013PhRvD..87l3526F,2015JCAP...11..022F,Fleury,Sanghai:2017yyn}, and of the related averaging issues~\cite{2017JCAP...03..062F,2016MNRAS.455.4518K,2015JCAP...07..040B}. Nevertheless, apart from Ref.~\cite{1981GReGr..13.1157D}, the Ricci-Weyl problem has never been discussed as the fundamental question it actually is: How can the laws of optics depend on the choice of the resolution of distribution of matter?

This Letter proposes a strong lensing formalism for weak lensing, which allows one to work beyond the infinitesimal light beam approximation. It provides a clear connection between the weak and strong lensing regimes, and solves the Ricci-Weyl dichotomy. Furthermore, it predicts that the weak lensing of finite light beams exhibits qualitative differences compared to the infinitesimal case; in particular, the Kaiser-Squires theorem~\cite{1993ApJ...404..441K} is violated by finite-size effects.

\section{The infinitesimal beam approximation}

Let us first be more specific about the underlying assumptions of the infinitesimal-beam approximation. From a physical perspective, it corresponds to assuming that the beam's cross-sectional diameter~$d$ is much smaller than all the characteristic geometrical lengths of the problem at hand. There are essentially three of them: first, the curvature radius of light's wave front, which is of the order of the angular distance~$D$ to the observer (where the beam converges), and which tells the typical distance over which the wave's amplitude varies appreciably; second, the typical spacetime curvature radius~$L_1\sim (R_{\mu\nu\rho\sigma}R^{\mu\nu\rho\sigma})^{-1/4}\sim 1/\sqrt{ G \varrho }$, driven by the mean energy density~$\varrho$ of the beam's neighborhood; third, the typical distance~$L_2 \sim L_1/\nabla L_1$ over which spacetime curvature varies appreciably, and corresponds to the typical inhomogeneity scale of the energy distribution. In cosmology, for example, a perturbation mode~$k$ of the density contrast has $L_2\sim 1/k$.

An infinitesimal beam satisfies $d\ll D, L_1, L_2$. Each of these inequalities has its proper physical interpretation:
\begin{enumerate}
\item\label{it:plane_parallel} $d\ll D$ (small angles) is the \emph{plane-parallel} approximation of lensing, or the flat-sky approximation in cosmology, also known as paraxial or Gaussian conditions in geometric optics. In the Sachs formalism, it ensures that the cross section of a light beam can be described within a unique screen space.
\item\label{it:weak_field} $d\ll L_1$ corresponds to the \emph{weak-field} regime, i.e., that spacetime's metric can be considered locally flat within the beam's cross section, allowing one to unambiguously define vectors, tensors, and distances at the beam's scale, including~$d$ itself.
\item\label{it:curvature_smoothness} $d\ll L_2$, finally, could be called the \emph{smooth curvature} approximation; it corresponds to the assumption that Riemann curvature can be considered constant (single-valued) within the beam's cross section, ensuring, in particular the validity of the geodesic deviation equation.
\end{enumerate}
While both~\ref{it:plane_parallel} and \ref{it:weak_field} are easily satisfied---and will be considered so throughout this Letter,---\ref{it:curvature_smoothness} is, in fact, always wrong when the light beam encounters some matter. Indeed, the microscopic, or even mesoscopic, structure of matter involves length scales which are clearly much smaller than any astronomical light beam. The purpose of this work is, thus, to understand how~\ref{it:curvature_smoothness} can be considered \emph{effectively} valid or not.

\section{Strong lensing formalism for weak lensing}

Consider an extended source at a distance~$D$ from the observer. For simplicity, we assume a background Minkowski geometry, so that distances are unambiguous, but our results are easily transposed to a cosmological context. Suppose that there are $N$ pointlike lenses distributed about the line of sight, which are allowed to be enclosed by the light beam corresponding to the observation of the source. This clearly violates assumption~\ref{it:curvature_smoothness}, so that the beam cannot be considered infinitesimal, and hence, the Sachs formalism cannot be applied. We choose to proceed with a strong-lensing-like approach, that is, somehow in the opposite way to the recent roulette approach~\cite{2016CQGra..33pLT01C,2016CQGra..33x5003C}.

\begin{figure}[h!]
\centering
\includegraphics[width=\columnwidth]{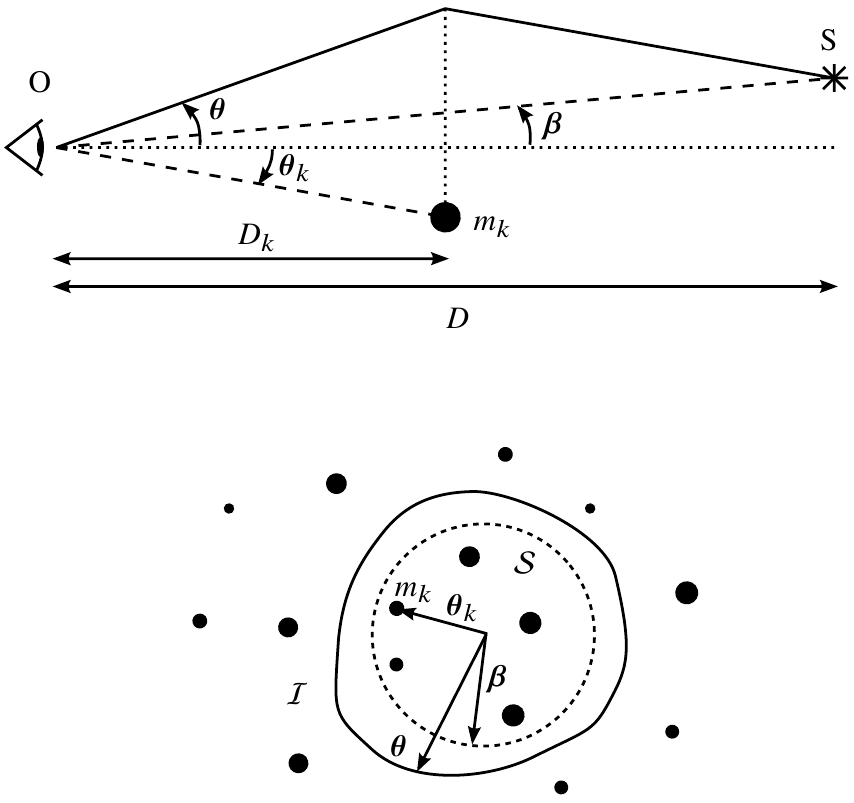}
\caption{Top panel: Lensing of a single point source $S$. Bottom panel: Contours of the source $\mathcal{S}$ and image $\mathcal{I}$, as seen by the observer.}
\label{fig:lensing}
\end{figure}

In the weak-field regime and plane-parallel approximation, each point of the extended source is mapped to its image through the lens equation~\cite{1992grle.book.....S}
\begin{equation}
\vect{\beta} = \vect{\theta} 
- \sum_{k=1}^N \eps_k^2 \, \frac{\vect{\theta}-\vect{\theta}_k}{\abs{\vect\theta-\vect{\theta}_k}^2},
\end{equation}
where~$\vect{\beta},\vect{\theta}$, and $\vect{\theta}_k$ denote, respectively, the angular positions, on the observer's celestial sphere, of the unlensed source, image, and lenses, as illustrated in Fig.~\ref{fig:lensing}. Besides, $\eps_k$ denotes the Einstein radius of the $k$th lens, with $\eps_k^2\define 4G m_k (D-D_k)/(D_k D)$, where $m_k$ is its mass and $D_k$ its distance to the observer.
The \emph{weak-lensing regime} (stricter than the weak-field regime) applies if the physical size of the lenses is much bigger than their Einstein radii, so that: (i) there is only one image per point source, and (ii) $\eps_k/\abs{\vect{\theta}-\vect{\theta}_k}\ll 1$. The Einstein radii~$\eps_k$ will thus be considered as small numbers in what follows.

Because they are two-dimensional vectors, $\vect{\beta},\vect{\theta}$, and $\vect{\theta}_k$ can be represented by complex numbers~$s, z$, and $w_k$, respectively, in terms of which the lens equation now reads
\begin{equation}\label{eq:lens_equation_complex}
s = z - \sum_{k=1}^N \frac{\eps_k^2}{z^* - w_k^*},
\end{equation}
where a star denotes complex conjugation. In the weak-lensing regime, this equation can be inverted order by order in $\eps_k^2$ as $z(s)=s+\delta^{(2)}z(s)+\mathcal{O}(\eps^4)$, where $\delta^{(2)}z(s)\define \sum_{k=1}^N \frac{\eps_k^2}{s^* - w_k^*}$. For the remainder of this Letter, we will work at order 2 in $\eps$ and drop the (2) superscript.

\section{From Weyl to Ricci}

We first investigate the lensing magnification of the extended source. We call~$\source$ its contour in the complex plane, which is mapped to the image contour~$\image$ via Eq.~\eqref{eq:lens_equation_complex}. The angular size of the image is, by definition,
\begin{equation}
\Omega = \frac{1}{2\i} \int_{\mathcal{I}} z^* \dd z .
\end{equation}
Substituting three times the lens equation, we get
\begin{multline}
\Omega = 
\frac{1}{2\i} \int_{\source} s^* \dd s
+ \frac{1}{2\i} \sum_{k=1}^N \eps_k^2 \int_{\image} \frac{\dd z}{z-w_k}\\
- \frac{1}{2\i} \sum_{k=1}^N \eps_k^2 
	\pac{ \int_{\image} \frac{z\dd z}{(z-w_k)^2} }^*
+ \mathcal{O}(\eps^4).
\end{multline}
The first term on the right-hand side is the unlensed size of the source~$\Omega\e{S}$, while the two other complex integrals can be computed using the residue theorem. Interestingly, only the lenses which are enclosed by the light beam, i.e. such that~$w_k$ lies in the region delimited by~$\mathcal{I}$, contribute to the angular size at that order. Computing the associated residues and replacing $\eps_k$ by its expression then yields
\begin{equation}\label{eq:Omega}
\Omega = \Omega\e{S} + \sum_{k\in\image} \frac{8\pi G m_k (D-D_k)}{D D_k}
+ \mathcal{O}(\eps^4),
\end{equation}
where $k\in\image$ means that we sum over lenses~$k$ which are enclosed by the image~$\image$. 

It is instructive to compare this result with the case of infinitesimal beams. At lowest order, Sachs equations imply that the convergence~$\kappa =(\Omega-\Omega\e{S})/2\Omega\e{S}$ is driven by Ricci curvature according to
\begin{equation}\label{eq:kappa}
\kappa = \frac{1}{2}\int_0^\lambda \frac{\lambda'(\lambda-\lambda')}{\lambda} \, (R_{\mu\nu} k^\mu k^\nu) \; \dd \lambda',
\end{equation}
where $\lambda$ is the affine distance and $k^\mu$ the associated wave four-vector. Substituting Einstein's equation in Eq.~\eqref{eq:kappa}, we find that it exactly matches Eq.~\eqref{eq:Omega} if the matter density on the line of sight is taken to be
\begin{equation}\label{eq:rho}
\varrho(r) = \sum_{k\in\image}^N \frac{m_k}{A_k} \, \delta(\lambda-\lambda_k),
\end{equation}
where $A_k=D_k^2\Omega$ is the physical area of the beam at $\lambda_k=D_k$.

This result generalizes Ref.~\cite{1981GReGr..13.1157D}. It shows that the area of a finite light beam, whose contour experiences Weyl curvature only, propagates like the area of an infinitesimal beam experiencing Ricci curvature only. Furthermore, this effective Ricci curvature is equal to the average Ricci curvature encountered inside the finite beam. In other words, the effect of a point mass~$m_k$ enclosed by the beam is identical to the effect of a homogeneous distribution of mass with surface density $m_k/A_k$, whatever its transverse position across the finite beam. This shows the ability of light beams to smooth out the matter distribution they enclose.

This property can be understood as a consequence of Gauss' theorem, which emphasizes the special status of the $1/r^2$ behavior of gravitation. Similarly to cosmological dynamics~\cite{Fleury:2016tsz}, we expect the discrete-to-continuous transition to be affected in a nontrivial way for alternative theories of gravitation.

Note, finally, that in the above we adopted the geometric notion of convergence, as opposed to its energetic counterpart~$\kappa_I=(I-I\e{S})/2 I\e{S}$, defined as the relative enhancement of luminous intensity. Both notions are known to be equivalent for infinitesimal beams, thanks to Etherington's reciprocity law~\cite{1933PMag...15..761E}, if the photon number is conserved. This is expected to hold here because we have assumed the lenses enclosed by the beam to be transparent and subcritical. But if the lenses were either partially opaque or smaller than their Einstein radii, this would effectively punch holes in the image, which could then appear larger ($\kappa>0$) but fainter ($\kappa_I<0$).

\section{Morphology of a finite beam}

While only the lenses that are enclosed by the light beam affect its area (at lowest order), all the lenses turn out to contribute to its distortions. Standard weak-lensing analyses precisely consist in measuring such distortions, in particular, in the apparent shape of lensed galaxies. The ellipticity of a galaxy is usually measured from the image quadrupole~\cite{2001PhR...340..291B}
\begin{equation}\label{eq:def_quadrupole}
\quadrupole_{ab} 
= \frac{\int_{\image} W[I(\vect{\theta})] \, \theta_a \theta_b \; \dd^2\vect{\theta}}
			{\int_{\image} W[I(\vect{\theta})] \; \dd^2\vect{\theta}},
\end{equation}
where $I(\vect{\theta})$ denotes the luminous intensity at the observed position~$\vect{\theta}$ and $W(I)$ is a weighting function. The complex ellipticity of the image~$\image$ is then defined as
\begin{equation}
\chi = \frac{\quadrupole_{11}-\quadrupole_{22} + 2\i \quadrupole_{12}}{\quadrupole_{11}+\quadrupole_{22}} .
\end{equation}

In the standard framework, based on the infinitesimal-beam approximation, the lensing contribution to $\vect{\quadrupole}$ is quantified by the amplification matrix~$\vect{\amplification}$ via~$\vect{\quadrupole}=\transpose{\vect{\amplification}}\vect{\quadrupole}_{\source}\vect{\amplification}$, where $\vect{\quadrupole}_{\source}$ is the quadrupole of the source. At lowest order in lensing, it implies that the complex ellipticity reads
\begin{equation}\label{eq:shear}
\chi = \chi_{\source} + 2 \gamma,
\end{equation}
where $\gamma$ is the complex shear due to gravitational lensing.

The above reasoning, in particular the relation between observed and intrinsic quadrupoles, is valid only if the amplification matrix can be considered \emph{homogeneous} across the image. This does not hold for a finite beam, where on the contrary some significant variations of~$\vect{\amplification}$ can occur within the image. However, we can still use Eq.~\eqref{eq:shear} in order to define the analog of shear in our finite-beam case. For a quasicircular source with average angular radius~$\beta$, assuming $W=1$ in the definition~\eqref{eq:def_quadrupole} of $\vect{\quadrupole}$, and at lowest order in $\eps$, we then find
\begin{equation}\label{eq:shear_Fourier}
\gamma = \frac{2\delta\theta_{-2}}{\beta}
\end{equation}
where we introduced the Fourier mode $\ell=-2$ of~$\image$, seen as a polar curve~$\theta(\psi)\define|\vect{\theta}|(\psi)$, with
\begin{equation}
\theta(\psi) = \sum_{\ell\in\mathbb{Z}} \theta_\ell \ex{\i\ell\psi}, \qquad
\theta_\ell \define \frac{1}{2\pi} \int_0^{2\pi} \theta(\psi) \ex{-\i\ell\psi} \; \dd\psi ;
\end{equation}
while the notation~$\delta \theta_\ell$ in Eq.~\eqref{eq:shear_Fourier} means that we isolated the contribution to $\theta_\ell$ due to lensing only (subtracting the source). Note, by the way, that $\kappa=\delta \theta_0/(4\beta)$.

\begin{figure*}[ht]
\centering
\includegraphics[width=0.19\textwidth]{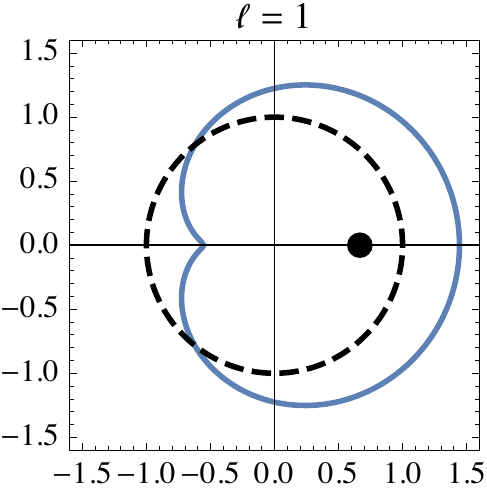}
\includegraphics[width=0.19\textwidth]{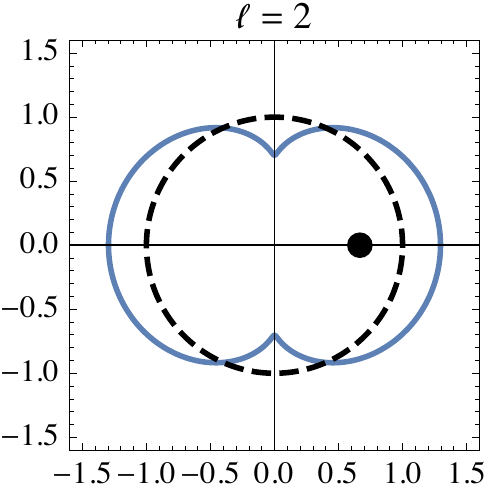}
\includegraphics[width=0.19\textwidth]{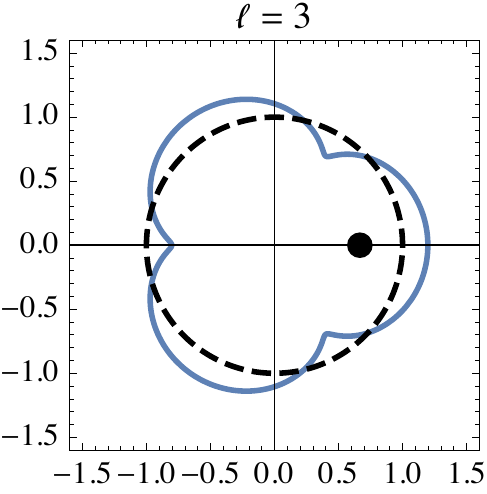}
\includegraphics[width=0.19\textwidth]{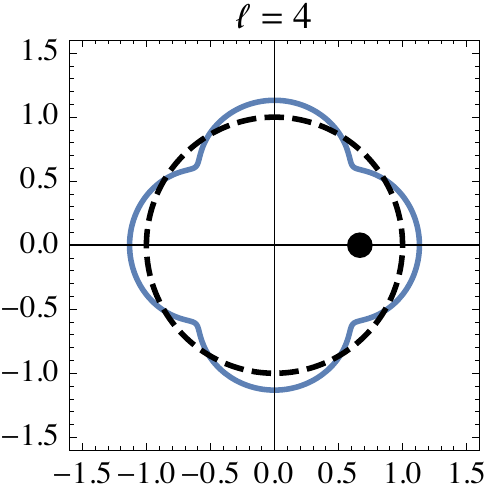}
\includegraphics[width=0.19\textwidth]{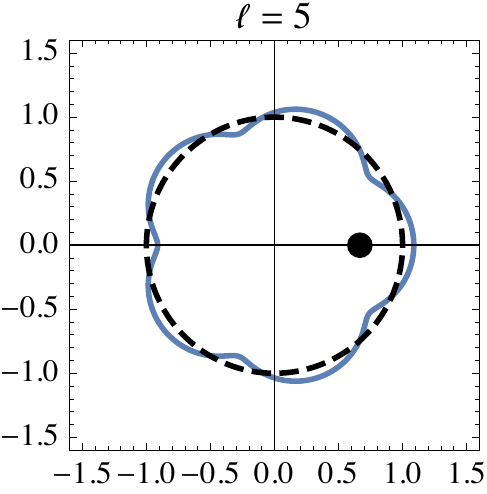}\\
\includegraphics[width=0.19\textwidth]{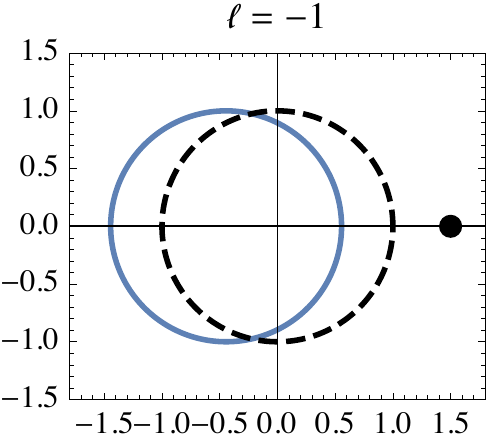}
\includegraphics[width=0.19\textwidth]{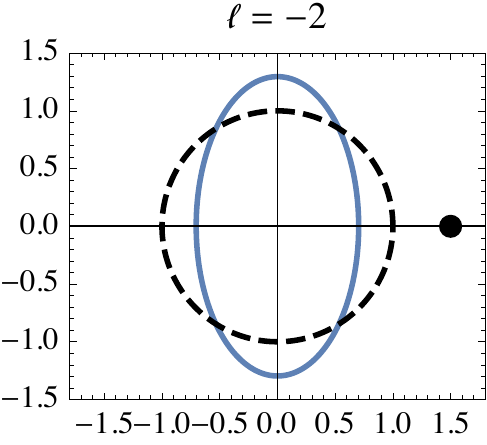}
\includegraphics[width=0.19\textwidth]{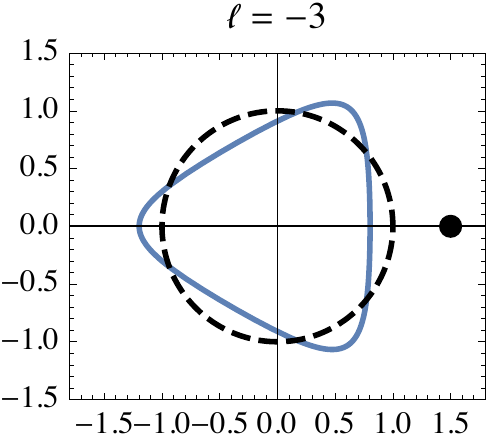}
\includegraphics[width=0.19\textwidth]{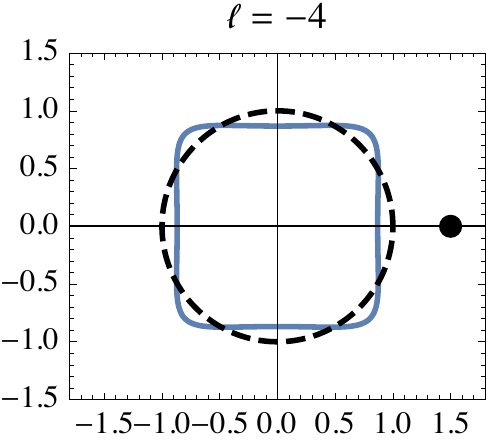}
\includegraphics[width=0.19\textwidth]{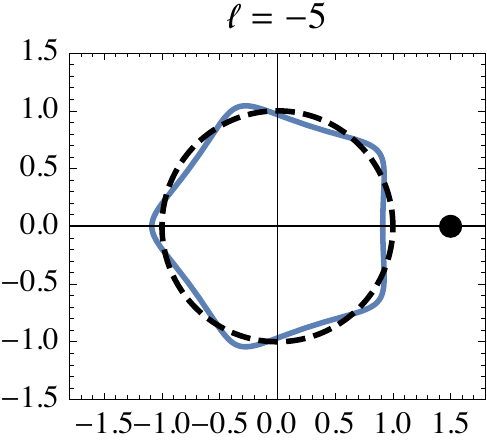}
\caption{Lowest Fourier modes of an image shape, generated from a circular source. The axes are normalized by the angular radius~$\beta$ of the source~$s=\beta\ex{\i\ph}$ (dashed lines). Solid lines indicate the sum $s+\delta z_\ell \ex{-\i(\ell+1)\ph}$. \textit{Top panel}: The lens is at~$w=2\beta/3$, inside the beam, and generates $\ell\geq 0$ modes only. \textit{Bottom panel}:  The lens is at~$w=3\beta/2$, out of the beam, and generates $\ell<0$ modes only. We have taken~$\eps^2=\beta^2/2 \not\ll |w-s|^2$ in order to visually enhance the effects.}
\label{fig:Fourier_modes}
\end{figure*}

Contrary to the infinitesimal case, the impact of gravitational lensing on the morphology of a finite image is distributed over an infinity of Fourier modes, of which shear captures only a tiny part. Let us be more explicit and calculate the $\delta \theta_\ell$ generated by an arbitrary distribution of lenses. With our complex formalism, we first find that
\begin{equation}\label{eq:delta_theta_ell}
\delta\theta_\ell = \frac{\delta z_\ell + (\delta z_{-\ell})^*}{2},
\end{equation}
where we defined the Fourier modes of the complex representation~$z$ of the image as
\begin{equation}
z_\ell \define \frac{1}{2\pi} \int_0^{2\pi} z[s(\ph)] \ex{-\i(\ell+1)\ph} \dd\ph.
\end{equation}
Note that integration is performed, here, with respect to the polar angle~$\ph$ \emph{of the source}, i.e., such that~$s=\abs{s}\ex{\i\ph}$, and not with respect to the polar angle~$\psi$ of its image~$z(s)=\abs{z}\ex{\i\psi}$. This difference is worth noticing, because it implies that $z_\ell$ is not an observable, unlike~$\theta_\ell$. Indeed, observation gives direct access only to images, not to sources. The quantities~$\delta z_\ell$, however, enjoy a very elegant expression. Using Eq.~\eqref{eq:lens_equation_complex} and the residue theorem yields
\begin{align}
\delta z_{\ell\geq 0} &= \frac{1}{\beta}  \sum_{k\in\image} \eps_k^2 \pa{\frac{w_k^*}{\beta}}^\ell, \\
\delta z_{\ell<0} &= -\frac{1}{\beta} \sum_{k\not\in\image} \eps_k^2 \pa{\frac{w_k^*}{\beta}}^\ell .
\end{align}
The positive Fourier modes are sourced only by interior lenses, while negative modes are sourced by exterior lenses. For illustration, the effect of the first ten distortion modes of a circular source is depicted in Fig.~\ref{fig:Fourier_modes}. This result is a special case of the normal modes (roulettes) calculated in Refs.~\cite{2016CQGra..33pLT01C,2016CQGra..33x5003C}, for a general spacetime geometry.

An important consequence of Eqs.~\eqref{eq:shear_Fourier} and \eqref{eq:delta_theta_ell} is that interior lenses contribute to the observed shear~$\gamma$, through $\delta z_2^*$. This is a key difference with the infinitesimal-beam case, where only exterior lenses are able to shear light beams. It is also remarkable that the effect of the latter remains rigorously unchanged, even for very large beams, which can cover regions across which the tidal field of the lens can vary appreciably.

\section{Finite-beam corrections in weak lensing}

We now investigate how the contribution of interior lenses to the observed shear could change qualitatively the interpretation of weak-lensing surveys. In the standard lore, cosmic shear measurements give access to the large-scale structure thanks to the Kaiser-Squires relation~\cite{1993ApJ...404..441K}, which states that shear~$\gamma$ and convergence~$\kappa$ have the same power spectrum, the latter being directly related to the matter power spectrum. As a consequence, $\kappa$ and $\gamma$ must have the same statistical variance,
\begin{equation}\label{eq:Kaiser-Squires_infinitesimal}
\ev[2]{\kappa^2}-\ev[2]{\kappa}^2 = \ev[2]{|\gamma|^2}
\qquad \text{(infinitesimal beam).}
\end{equation}
Note the presence of a nonvanishing~$\ev{\kappa}$, due to the fact that we are working with a Minkowski background. In the standard lore, this quantity is contained in the Friedmann-Lema\^{i}tre-Robertson-Walker background. Equation~\eqref{eq:Kaiser-Squires_infinitesimal} turns out to be generically violated when finite-beam effects are accounted for.

Consider a simple static model where the Universe is randomly filled with point lenses, with random masses~$m$, whose distribution is statistically homogeneous, with mean number density~$\bar{n}$. The convergence is found to read
\begin{equation}
\ev[2]{\kappa^2}-\ev[2]{\kappa}^2 = \frac{4\pi}{3\beta^2} \, \bar{n}\ev[2]{r\e{S}^2} D,
\end{equation}
where $r\e{S}=2Gm$ denotes the Schwarzschild radius of a lens with mass~$m$. As expected, the variance of $\kappa$ vanishes in the continuous limit, where the Universe becomes strictly homogeneous ($\ev{m}\rightarrow 0$, $\bar{n}\rightarrow\infty$, $\bar{n}\ev{m}=\bar{\varrho}$). However, it diverges as the size of the beam, $\beta$, goes to zero. This limit is, however, ill-defined in our framework, because for a very small beam one cannot neglect the lenses' Einstein radii~$\eps^2$ any more.

The variance of the other lensing Fourier modes reads
\begin{equation}
\ev[2]{|\delta z_\ell|^2} = \frac{4\pi}{3} \frac{\bar{n}\ev[2]{r\e{S}^2}D}{|\ell+1|}.
\end{equation}
for $\ell\not= -1,0$. The case $\ell=-1$ diverges, which is most probably an artifact of the plane-parallel approximation~\cite{2008MNRAS.386..230W}. This mode corresponds to the global (unobservable) displacement of the image and, hence, does not affect its shape. Equations~\eqref{eq:shear_Fourier} and \eqref{eq:delta_theta_ell} then yield
\begin{equation}\label{eq:Kaiser-Squires_finite}
\ev[2]{\kappa^2}-\ev[2]{\kappa}^2 = \frac{3}{4} \ev[2]{|\gamma|^2}
\qquad \text{(finite beam)},
\end{equation}
which displays a significant discrepancy with the infinitesimal-beam case~\eqref{eq:Kaiser-Squires_infinitesimal}. This is due to the presence of the $\ev[1]{|\delta z_2|^2}$ term in Eq.~\eqref{eq:shear_Fourier}, related to interior lenses, whose contribution to $\gamma$ turns out to be statistically comparable to the standard $\ev[1]{|\delta z_2|^2}$ due to exterior lenses. The net result is the enhancement by a factor $4/3$ of the actual shear variance, whence the factor~$3/4$ compared to the Kaiser-Squires relation~\eqref{eq:Kaiser-Squires_finite}.

\section{Conclusion}

In this Letter, we analyzed the properties of weak gravitational lensing beyond the infinitesimal-beam approximation. We addressed the Ricci-Weyl problem by showing that light beams are smoothing out the distribution of matter they enclose, at the scale of the beam's cross section. While only the lenses which are enclosed by the beam contribute to the convergence, at lowest order, any lens contribute to its distortions. Such distortions of the beam's morphology were decomposed over Fourier modes, elegantly expressed in terms of the position and mass of the lenses. 

In particular, the standard weak-lensing shear was found to involve not only lenses out of the beam, but also interior lenses. In a Universe filled with point lenses, this property statistically enhances the dispersion of shear~$\ev[1]{|\gamma|^2}$ by a factor or $4/3$, implying a qualitative violation of the Kaiser-Squires theorem. This finite-beam correction, however, may be overestimated by the simplicity of our model for matter distribution. Note that current ray-tracing techniques in $N$-body simulations are unable to evaluate this effect, because they all rely on the standard infinitesimal-beam formalism in order to compute the weak-lensing convergence and shear~\cite{2000ApJ...530..547J,2016MNRAS.461..209G,2016JCAP...05..001B}. Going beyond this requires the reconstruction of light beams from a large number of light rays, similarly to how one generates magnification maps in microlensing~\cite{1986A&A...166...36K,microlensing_teraflop}.

Shall a significant correction hold for more realistic models for the distribution of matter, it would have profound consequences on both theoretical and observational aspects of cosmology. On the theory side, the relation~\eqref{eq:Kaiser-Squires_infinitesimal} between convergence and shear is essential to the calculations of the bias of the luminosity-redshift relation by cosmological perturbations~\cite{2015JCAP...07..040B,2017JCAP...03..062F}. On the observational side, in the analysis of weak-lensing surveys, the Kaiser-Squires theorem is essential for relating the shear power spectrum to the matter power spectrum. An excess of shear means that we would overestimate the combination of $\Omega\e{m}$ and $\sigma_8$ inferred from cosmic shear surveys, worsening the tension with \textsl{Planck} measurements~\cite{2017MNRAS.471.4412K}. The necessary evaluation of the finite-beam correction to the full shear two-point correlation function will be addressed in a followup article, as well as the possible corrections to Etherington's reciprocity relation and its main consequence: the distance duality relation.


\section{Acknowledgements} We thank Cyril Pitrou, George Ellis, Ruth Durrer, and Chris Clarkson for discussions. P. F. and J.-P. U. thank UCT for hospitality during the early phases of this work. P. F. acknowledges support by the Swiss National Science Foundation. The work of J.-P. U. is made in the ILP LABEX (under reference ANR-10-LABX-63) was supported by French state funds managed by the ANR within the Investissements d'Avenir program under reference ANR-11-IDEX-0004-02. J. L.'s work is supported by the National Research Foundation (South Africa).


\bibliography{bibliography_Ricci-Weyl}

\end{document}